\documentclass[10pt]{article}
\usepackage{arxiv}

  \usepackage{cite}

  \usepackage[pdftex]{graphicx}
  \graphicspath{{../IEEEtran/}}
\usepackage{url}
\usepackage[x11names]{xcolor}

\usepackage[colorinlistoftodos,prependcaption,textsize=tiny]{todonotes}
\usepackage{changes}

\begin{document}
%
\title{An Adaptable Approach for Successful SIEM Adoption in Companies}
%
%

\author{Maximilian Rosenberg, Bettina Schneider, Christopher Scherb and Petra Maria Asprion}


\maketitle
\begin{abstract}
  In corporations around the world, the topic of cybersecurity and information security is becoming increasingly important as the number of cyberattacks on themselves continues to grow. Nowadays, it is no longer just a matter of protecting against cyberattacks, but rather of detecting such attacks at an early stage and responding accordingly. There is currently no generic methodological approach for the implementation of Security Information and Event Management (SIEM) systems that takes academic aspects into account and can be applied independently of the product or developers of the systems. Applying Hevner’s design science research approach, the goal of this paper is to develop a holistic procedure model for implementing respective SIEM systems in corporations. According to the study during the validation phase, the procedure model was verified to be applicable. As desire for future research, the procedure model should be applied in various implementation projects in different enterprises to analyze its applicability and completeness.
\end{abstract}





%

\section{Introduction}\label{sec:introduction}

Cybersecurity is a major concern and a rising threat for corporations. While the focus in the past was on preventing attack from the outset ("protection"), measures are increasingly being taken to detect and respond to attacks in an early stage ("detect" and "respond"). It has to be accepted that cyber attacks are the rule - not an exception - and they have to be managed holistically~\cite{bygrave_cyber_2022}. This view of cybersecurity is also consistent with the recommendations of the National Institute of Standards and Technology (NIST), one of the leading US organization in the cybersecurity field~\cite{national_institute_of_standards_and_technology_framework_2018}. One way to track security-releated activities in the IT systems or the entire IT landscape of a company is with the help of a Security Information and Event Management (SIEM) system. SIEM systems collect, store, and analyze security-related logs that provide information related to information, network, and data security as well as regulatory compliance~\cite{chuvakin_complete_2021}.

Only when properly implemented, SIEM systems can unfold their value in corporations in terms of security. The implementation is a resource-intense and demanding procedure. If a SIEM system is not implemented correctly, it can result in a corporation having an expensive system in place which does not achieve what they initially expected from the system~\cite{mokalled_guidelines_2019}. Additionally, the system may ultimately not be used because it is not a suitable configuration for the company. It is essential to choose the right system and to customize it in the right way. As skilled professionals in cybersecurity are scarce, it is proper to support the employees in this procedure~\cite{cybersecurity_workforce_isc_nodate}. While generic approaches exist how to implement enterprise system, there is no methodological procedure for the implementation of SIEM systems in the corporate context. The implementation of any information system - whereas SIEM system can be seen as an information system - is a critical process that has already been extensively researched due to its importance. Researchers have developed various models in their studies on how such systems can be implemented~\cite{huang_comprehensive_2016}. Consequently, there are many procedure models that depict the implementation of enterprise resource planning (ERP) systems (see e.g.~\cite{kurbel_erp_2013, kronbichler_comparison_2009}).

A literature search conducted on the platforms Google Scholar, IEEE Xplorer, ResearchGate, Scopus and Swisscovery for the keywords "SIEM", "implementation", "deployment", and "procedure model" has shown that there are procedures from the developers themselves, but no product-independent procedure model from academia\footnote[1]{Section 3 describes in more detail what the results of this literature search revealed.}. An analysis of existing process models showed the following gaps: 1) no SIEM implementation model could be found that depicted the phases "evaluation", "deployment" and "operation". 2) procedure models of SIEM product vendors provide inputs for evaluating a system, however a more neutral view would be desirable. 3)  A model for turning the implemented system over to operations after successful deployment to operations could not be found during the research. This paper will close this gap and elaborate the development of a methodology how to implement SIEM systems in corporation context.

The remainder of this paper is structured as follows. In section 2, the methodology of this study elaborated. Section 3 explains the core of the given problem and defines the requirements for the procedure model. In Section 4, existing generic procedure models and security frameworks are analyzed to identify elements that could be used in our new model. Section 5 explains how the SIEM-related procedure model was developed based on the research phases. The procedure model was presented to cybersecurity experts for validation, additionally certain parts of the developed artifact were applied for validation in a company. These findings and the final procedure model are presented in Section 6. Finally, in section 7,conclusions are drawn, and an outlook regarding the procedure model is described.

\section{Methodology}
To ensure the quality of this paper, the problem-solving Design Science Research (DSR) approach by Alan Hevner~\cite{hevner_design_2004} was used. This approach aims to generate knowledge through the creation and design of artifacts. According to Hevner, an artifact can be software, a method, model, or concept. Further processing of an artifact can in turn generate further knowledge, which can be used to generate further artifacts.

\begin{figure}[!htb]
  \centering
  \includegraphics[width=88mm]{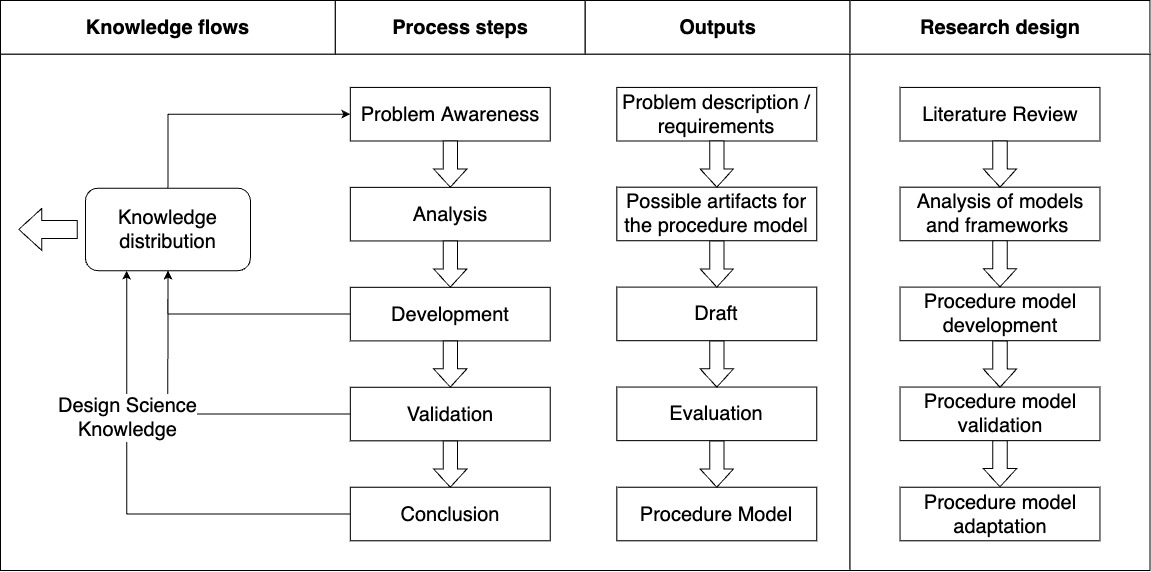}
  \caption{Design Science Research Design according to Kuechler and Vaishnavi~\cite{kuechler_theory_2008}.} 
  \label{Figure 1}
\end{figure}

A useful DSR starts with the identification of an actual problem, which in this paper is the lack of a procedure model for the implementation of SIEM systems in companies. Figure \ref{Figure 1} shows the five process steps that were performed to develop the procedure model.

\textit{Problem Awareness}: The first DSR process step was performed to create awareness of why a procedure model is needed for the implementation of SIEM systems. Awareness of the problem was raised by conducting a literature review on log management, log analysis and the current state of SIEM systems. In addition, the basic functionalities of SIEM systems were investigated (see section 3).

\textit{Analysis}: In the second DSR process step, existing procedure models in related fields were analyzed. Security frameworks were also examined that could be of interest to companies and could provide support in the implementation of a SIEM system. In addition to the theoretical analysis, a case study was conducted at a Swiss retail company. A SIEM system was evaluated and implemented for the company. Due to this, practical experience could also be incorporated into the new procedure model (see section 4).

\textit{Development}: After the awareness of the problem was created and an analysis was performed, the development of the procedure model followed. For this, the previously acquired knowledge was combined based on the theoretical and practical findings that were gathered in the analysis phase (see section 5).

\textit{Validation}: Interviews were conducted with cybersecurity experts to validate the new procedure model. They were presented with the procedure model and asked for feedback on the evaluation of the model. Based on these interviews, further insights could be gained, which in turn could be incorporated into the model developed up to that point \footnote[1]{Usually, the fourth DSR process step is called “Evaluation”. Since a phase is referred to as “Evaluation” or “Evaluation phase” in the procedure model, this DSR process step is referred to as “Validation” in this paper. This provides readers with a clear separation between the evaluation process step in the DSR and the evaluation phase within the procedure model.} (see section 6).

\textit{Conclusion}: In the last DSR process step, the procedure model was revised and finalized based on the inputs and findings of the validation step. The result of the various DSR process steps is an applicable procedure model for the implementation of SIEM systems in companies (see section 7).

\section{Problem Awareness}
In this section, the problem description and the requirements for a SIEM procedure model are elaborated. It starts by looking at the background of log management systems as they are a relevant base for SIEM systems. The result of this section is the problem description and the requirements for a new procedure model for the implementation of SIEM systems in companies.

\subsection{From Log Management to SIEM Systems}
Logs are records of events that occur in an organization's systems, networks, or applications. Each entry in a log contains information about a specific event that took place. While originally, logs were used primarily for troubleshooting, nowadays they serve other purposes such as recording user interactions, performance, and identifying malicious activity, too. An example of a security event is the authentication of a user attempting to access a system. As threats to corporate networks and systems continue to grow, the need to manage and analyze logs has emerged to improve threat detection and fast incidence response~\cite{kent_guide_2006,sahoo_syslog_2019}.

Log management is the process of managing logs generated by various systems in an organization. According to Kent and Souppaya~\cite{kent_guide_2006}, this process involves the generation, transmission, storage, analysis and deletion of security logs. The generation of logs usually takes place on the host on which an application, software or service is operated. Based on the aforementioned process, these logs or the individual events of the logs would have to be transferred and stored. Therefore, companies use centralized log management systems that receive or retrieve and store the logs from the various hosts. Afterwards, the centrally stored logs should be analyzed, kept for a defined period of time, and deleted afterwards. The overall goal is to centrally manage and analyze different types of logs from different systems.

Log management systems offer various basic functionalities such as "Log Collection", "Log Filtering", "Log Reduction", "Log Parsing", "Log Normalization", "Log Rotation, "Log Compression", "Log Archival", "Event Correlation" and "Log Analysis" (Search and Reporting).

Log management systems have the strength of standard functionalities to collect, store, and rotate logs but they are falling short in identifying anomalies and creating timelines of events, users and assets which should be monitored from a SIEM. This is why today they are still a relevant base, but developed further to SIEM systems, discussed in the following section.

A SIEM system builds on the fundamentals of log management systems. While a log management system collects and stores all types of logs, a SIEM focuses on security-relevant logs that provide information related to network, and data security, as well as regulatory compliance. Consequently, the most obvious difference between SIEM and log management systems is the focus of SIEMs on security related logs and their analysis.~\cite{chuvakin_complete_2021}.

The concept and functionality of SIEMs is the combination of Security Information Management (SIM) and Security Event Management (SEM). SIM collects security-related logs for report generation. SEM analyzes these security-related logs in real time to build threat models, correlate events, and thus detects security incidents early. SIEMs are mainly used by Security Operations Centers (SOC), which aim to maintain and improve the security of an organization~\cite{kent_guide_2006,miloslavskaya_analysis_2018, vielberth_security_2021}.

Most SIEM systems have the previously mentioned functionalities of log management systems. In addition, SIEMs offer further context related functionalities such as the integration of user and entity behavior analysis (UEBA) or machine learning-based data analysis.~\cite{chuvakin_complete_2021,yelevin_use_2021,salitin_role_2018,exabeam_exabeam_2021,voigt_ueba_2018}. 

SIEMs are used in medium and large enterprises as a proactive step to monitor IT security incidents. There are different implementations of SIEMs from different vendors, but their basic functionalities and mechanisms are similar. Incoming events are analyzed using various rules or data models and compared with past events. Alerts can be triggered, which inform about a strange event, or a multitude of them~\cite{kent_guide_2006,salitin_role_2018,shahid_anomaly_2021, scherb2018resolution, grewe2018network}. 

A SIEM helps IT security analysts detect as well as prioritize potential security incidents. Implementing a SIEM can provide an enterprise with several benefits, but - if done wrong - it can also have some disadvantages. One advantage is that a SIEM allows an organization to analyze security-related logs in real time using data models and machine learning. Based on the real-time analysis, security teams can react to incidents immediately and initiate countermeasures if necessary. Among others, the following threats can be detected by SIEMs: \textit{Anomaly Detection}, \textit{DDoS Attacks}, \textit{Botnet Activities}, \textit{Intrusion Attempts}, \textit{Ransomware} and \textit{Data Theft}~\cite{gonzalez-granadillo_security_2021}.

\subsection{SIEM Implementation}
Acquiring a SIEM can be expensive and take up a lot of resources. This applies to both implementation and operation. If the alerts of a SIEM are not analyzed and processed, the introduction of a SIEM does not bring any added value to the company. Depending on the product and type of installation (on-premise or cloud), costs can vary. This is a disadvantage for many companies when implementing a SIEM system and has the consequence that SIEMs are mainly deployed in corporations~\cite{sukma_analysis_2019}.

SIEM systems are usually implemented to map specific use cases. For example, a corporation could use the SIEM for reporting and regulatory compliance, insider threats and threat hunting. Therefore, it is important that a corporation knows which goal they want to archive with the implementation of a SIEM system. Consequently, the model must include an evaluation phase.

The latest Gartner Magic Quadrant analysis by Kavanagh et. al.~\cite{kavanagh_magic_2021} found that the SIEM systems market grew from 3.55 billion in 2019 to 3.58 billion in 2020. The report indicates that new customers mainly want to implement cloud-based systems. The reason for this desire is that via the cloud-based approach, SIEM software deployment and implementation can be simplified compared to on-premise deployments. Further, this report indicates that corporations are re-evaluating their current SIEM vendors. Reasons for these re-evaluations include incomplete and failed deployments. In a survey related to SIEM services, many customers told Gartner they needed outside support for implementation. More companies indicated that internal resources and expertise are not sufficient to manage a SIEM, so interest in external SOCs is expected to grow. Therefore, the adoption approach, that is developed in context of this research, should include a deployment and operation phase in addition to the before mentioned evaluation phase.

\subsection{Requirements for SIEM Implementation Procedure}
The search for procedure models for the implementation of SIEM systems that include the phases "evaluation", "deployment" and "operations" did not yield to any results. Therefore, a search was conducted for procedure models that represent at least one of these three phases. 

The research for models, which can be used for the evaluation of SIEM products, delivered few results. However, these are not effective models that show the individual steps in the evaluation, but only criteria should be considered when selecting a SIEM. They were also mostly guides from SIEM vendors themselves, such as Exabeam~\cite{exabeam_exabeam_2021}, AlienVault~\cite{alienvault_siem_2022}, and Splunk~\cite{splunk_leitfaden_2018}. In addition to these three guides, a whitepaper by Filkins~\cite{filkins_evaluators_2018} - funded by SIEM vendor LogRythm5 - could be found. Another result of the search was a paper by Safarzadeh et al.~\cite{safarzadeh_novel_2019}, which shows a possible assessment methodology for SIEM products. The assessment is based on three dimensions, none of which include the business requirements for the system. In conclusion, none of these results provided a model that can be applied to the evaluation phase.

The search for procedure models for the deployment of a SIEM did also not yield to applicable results. The only result of the search was a best practice manual for the deployment of Azure Sentinel6~\cite{grigorof_azure_2021} and a conference paper~\cite{holik_deployment_2015}, which describes the deployment of a SIEM in a cloud infrastructure. The paper also described possible deployments based on specific SIEM products such as AlienVault OSSIM and QRadar. Like when searching for evaluation models for SIEM products, the guidance, albeit practical, is specific to one or more specific products. The goal of this paper is to develop a procedure model that can be used generally for the implementation of SIEM systems.

The search for models describing how to operate SIEM systems and the operational application after the initial deployment did not yield any results. 

This research has shown that there is currently no procedure model for the implementation of SIEM systems in companies which can be applied independently of developer and products and which includes the phases "evaluation", "deployment" and "operation". On the basis of these results, requirements were defined which the new procedure model for the implementation of SIEM systems must fulfill which are described in Table \ref{Table 1}. The requirements will be taken up in Sections 4 and 5 in order to show that the defined requirements are met.

\begin{table}[!htb]
  \centering
  \begin{tabular}{||l|p{6.7cm}||}
    \textbf{Req. No.} & \textbf{Req. Description} \\
    R1 & The model is based on elements of existing project management methods. \\
    R2 & The model covers the three phases "evaluation", "deployment" and "operation". \\
    R3 & The model can be applied regardless of manufacturer or product. \\
    R4 & The model is based on agile methods and approaches. \\
    R5 & The model includes references to other methods and/or security frameworks. \\
  \end{tabular}
  \caption{Procedure Model Requirements}
  \label{Table 1}
\end{table}

\section{Analysis}
In this section, classical procedure models and security frameworks are analyzed in order to identify elements that can be used for the implementation of a SIEM system. The result of this section are the derived components from the analysis of the procedures and security frameworks.

\subsection{Implementation Procedures}
\begin{figure}[!htb]
  \centering
  \includegraphics[width=88mm]{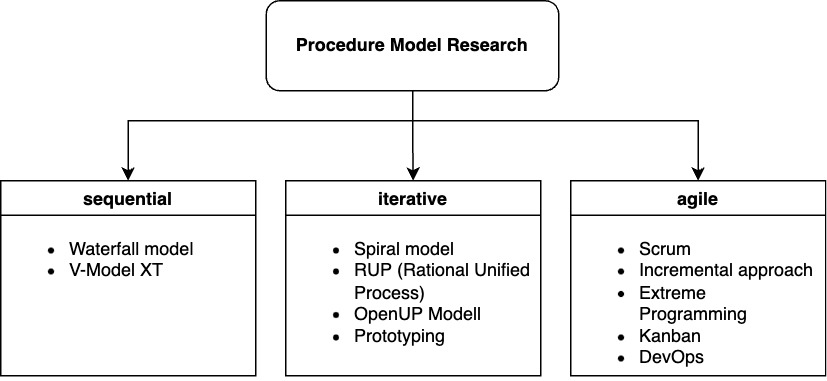}
  \caption{Procedure Models Research Result (based on [26 - 31])} 
  \label{Figure 2}
\end{figure}

One of the requirements for the procedure model is that it should be based on elements of existing models that are usually used (see R1 in Table \ref{Table 1}). The reason for this requirement is that the model should be easy to apply. Consequently, a literature research was conducted, which should provide an overview of the different procedure models. After the models were identified, they were divided into three categories (sequential, iterative or agile)~\cite{asprion2023agile}. This assignment was made so that it is clear which project management approaches are used in connection with software development or implementation in companies. The result of this research is shown in Figure \ref{Figure 2}~\cite{broy_vorgehensmodelle_2021,heil_vorgehensmodelle_2012,leyh_passende_2019,sandhaus_hybride_2014,schatten_vorgehensmodelle_2010,vivenzio_vorgehensmodelle_2013}.

\begin{table}[!htb]
  \centering
  \begin{tabular}{||l|p{6.7cm}||}
    \textbf{Crit. No.} & \textbf{Criteria Description} \\
    C1 & The model includes phases and/or activities. \\
    C2 & The model can be applied to two of the three phases (evaluation, deployment and operation). \\
    C3 & The model has agile characteristics or can be combined with agile methods. \\
    C4 & The model has return options or control mechanisms. \\
    C5 & The model is generally applicable. \\
  \end{tabular}
  \caption{Procedure Model Requirements}
  \label{Table 2}
\end{table}

Due to the large number of results, minimum criteria were defined, which can be seen in Table~\ref{Table 2}. They were determined based on the previous literature reviews. After the procedure models had been checked against these criteria, three procedure models were subsequently analyzed in more detail. In the following paragraphs, the origins of the three procedure models are shortly explained and their advantages and disadvantages are explained \footnotetext[1]{By purpose, we decided not to define the basics of each model as it is expected that reader either have some pre-knowlege or read the basics form the cited sources. All other models are not analyzed further in this paper.}. 

\textit{Waterfall model}: In 1970, Royce~\cite{royce_managing_1970} first described the model in his paper called "Managing the development of large software systems" by mapping the software development process into seven phases. The waterfall model is a phase-oriented process model. At the initialization of the project, different phases are defined, which are passed through until the completion of the project. These can vary depending on the type and nature of the project. The phases are run through sequentially from top to bottom during the project. Feedback makes it possible to "jump back" from the current phase to the previous phase in case certain results of the previous phases are needed, which were not or only partially delivered. In addition to the phases and feedback options, the waterfall model includes milestones. According to Gessler and Kaestner~\cite{gessler_projektphasen_2016}, these should be derived from the project goals. Typically, milestones represent events or results that must be achieved at the end of a phase. Accordingly, milestones serve as control elements to ensure goal- and result-oriented work in the project.

When a project is initialized, the phases are defined. In terms of design, the model offers flexibility, as the phases can be defined by users themselves. Consequently, the phases of the model can be customized based on the circumstances of each project. Once the phases are fixed, they cannot be changed during the project, which makes the model inflexible. Through the use of milestones, control and decision points can be established that facilitate the decision of whether a phase is complete and the project can move on to the subsequent phase. Therefore, the completion of one phase is necessary to move to the next phase, as the phases are passed sequentially one after the other. According to Schatten et. al.~\cite{schatten_vorgehensmodelle_2010}, the waterfall model should mainly be used for projects where the requirements can be clearly defined. The reason for this is that jumps back to past phases are only possible to a limited extent. In contrast to other procedure models, the work must be stopped during a return jump until the phase which had to be repeated has been completed again. This means that two phases cannot be carried out in parallel, since the prerequisite for the transition to a next phase is the completion of the previous one and they build on each other.

\textit{V-Model XT}: According to Kneuper~\cite{kneuper_geschichtliche_2018}, the concept of the V-model was introduced by the presentation of a sequential life cycle model by Boehm in 1979, which represented a V-shape. Verification and Validation (V\&V) were at the forefront of the model, as these steps were particularly important in the development of critical systems at that time. The model contains various procedure modules that can be individually combined to tailor the process model to the specific project. The documentation of the V-Modell contains various references to help users determine which modules can be used when. Among other things, it contains references to product types, roles, processes, and standards. These can be used to adapt the V-Modell to the circumstances of the project. The activities, which can be seen before the "V", are processed and run through sequentially, as in the waterfall model. Also after the "V" elements can be shown, which are run through in a sequential form after all iterations have been completed~\cite{broy_vorgehensmodelle_2021}. The elements within the "V" are run through in different iterations. This is an important component for software development, since new functionalities are developed during each iteration. As a result, the V- Model XT is a combination of a sequential and iterative approach~\cite{angermeier_v-modell_2019}.

Similar to the waterfall model, V-Modell XT considers the various procedure modules during the initialization of the project and adapts the model to the corresponding circumstances. This allows the model to be used in different forms and for different types of projects. The target group for this model is organizations that carry out software development projects~\cite{angermeier_v-modell_2019}. One advantage of V-Modell XT is the exact description of how and in which form the model can be applied in practice. The documentation provides a lot of information, hints and references on how the model can be used based on a circumstance. Similar to the waterfall model, V-Modell XT includes different phases, which can be defined at the beginning of the project and verified by delivering results. Nevertheless, like the waterfall model, this approach is strongly phase-oriented, which is why adjustments to the phases cannot be made during the project. In terms of tailoring, Kleuker~\cite{kleuker_vorgehensmodelle_2018} mentions that care should be taken when tailoring the model to avoid creating superfluous artifacts that do not add direct value to the project.

\textit{Scrum}: According to Volland~\cite{volland_scrum-framework_2021}, the term Scrum originates from the two organizational researchers Takeuchi and Nonaka~\cite{takeuchi_new_1986}. In the article "The New New Product Development Game" from 1986, they pointed out that the sequential and phase-oriented project management should be replaced. Instead, a unified and collaborative way of working should be used so that companies and their development teams can keep up with the rapid changes in the market. Scrum describes an iterative process, which enables the agile approach in software development. The main difference to the previous procedure models is that the focus in this approach is the prioritization of tasks. Likewise, the model can deal with changing requirements, which was not possible or only to a limited extent possible with the previously discussed procedure models. The model is mainly used in the field of software development. Scrum defines a general framework of how the activities in the project should be carried out. However, it does not prescribe how these activities are effectively carried out by the team. Scrum is based on the idea of an intelligent and self-organizing organization.

Scrum enables the team to work flexibly. The team organizes itself during the sprints and is not disturbed during this time. This allows the team to fully concentrate on the current tasks, which were selected at the Sprint Planning Meeting. Likewise, the development team itself plans which person takes over which tasks or subtasks. As a result, the strengths of each individual team member can be utilized, thus increasing efficiency. By prioritizing the requirements, which is carried out by the product owner in collaboration with the business, the requirements that are most important for the business can be implemented first. The use of Scrum is efficient for smaller teams, as communication during a Sprint and at the various meetings can be well coordinated. For larger teams, the process model can also be used, although the communication effort with regard to project management can be a challenge. Because the requirements are processed in subtasks during the sprints, results can already be delivered over a short period of time, which in turn can be presented to the business, the users or the customers. Scrum proves to be useful when results should be delivered within the shortest possible time. For example, Scrum could also be used for the development of a concept, whereby partial results are delivered again and again until finally the complete concept has been developed~\cite{schatten_vorgehensmodelle_2010,volland_scrum-framework_2021,lucht_theorie_2019}.

\subsection{Security Frameworks}
Another requirement for the new procedure model is that it includes references to methods and security frameworks (see R5 in Table \ref{Table 1}). These references should help users to apply the right methods and, most importantly, to consider well-known security frameworks in a SIEM project. Based on a research, three cybersecurity frameworks were selected, which are generally known in the cybersecurity field. These are the cybersecurity frameworks from NIST~\cite{nist_cybersecurity_2013}, the ISO/IEC 2700112 standard from the International Organization for Standardization (ISO)~\cite{iso_isoiec_nodate}, and the MITRE Att\&ck Framework from the MITRE Corporation~\cite{mitre_corporation_mitre_nodate}. All three frameworks originate from recognized authorities or organizations that are used in companies worldwide.

\subsubsection{NIST Cybersecurity Framework (CSF)}
The NIST CSF helps organizations manage risk with respect to cybersecurity risks. With this framework, NIST takes a risk-based approach to cybersecurity risk management. This CSF supports companies in the process of identifying, assessing and responding to risks. According to the framework, companies should know the probabilities of an event occurring and the possible resulting effects. With the help of the framework, companies can describe their risk tolerance. As a result, they can determine their level of risk, which is acceptable for achieving the company's goals. By knowing the risk tolerance, an organization can prioritize cybersecurity events and make decisions based on this information to faster fight critical threats~\cite{national_institute_of_standards_and_technology_framework_2018}.

\textit{Suitability and possible applications for SIEM implementation projects}: In the context of SIEMs, it makes sense to take a closer look at the "Detect" function. This function in turn contains various categories. One example of this is the category "Anomalies and Events". This category is about detecting anomalous activities and understanding the potential impact of these. Within this category, there are additional subcategories that provide more specific aspects of anomaly detection. The subcategories address different areas such as event detection and analysis (DE.AE-2) and event collection and correlation (DE.AE-3).

If the NIST CSF is already in use in an organization, the introduction of a SIEM can help to cover categories or various subcategories. However, the NIST CSF can also be a support for companies that do not currently use the framework. The framework provides many references in the individual subcategories to other frameworks and standards that could be considered when implementing a SIEM. The NIST CSF provides companies with information on how to improve risk management and the handling of cybersecurity risks from a strategic perspective.

\subsubsection{ISO/IEC 27001 Information Security Management}
The International Organization for Standardization (ISO) is an independent and non-governmental organization. A total of 167 national standards bodies count as its members. The goal of the organization is to bring experts together so that knowledge can be exchanged and thus international standards can be developed. The International Electrotechnical Commission (IEC) is a non-profit organization (NPO) that supports international trade in electrical and electronic goods. Like the ISO, the IEC consists of various members from over 170 countries. Together, the ISO and IEC form a committee that has already developed various standards in the field of information technology.

ISO/IEC 27001:2013 defines requirements for an information security management system (ISMS). It should be mentioned here that ISO/IEC 27001 is a standard and not a framework, which is why ISO/IEC 27001:2013 is referred to as a standard in the further course. This standard contains requirements for the establishment, implementation, maintenance and improvement of an ISMS. Along with these requirements, it provides information on how information security risks could be assessed and addressed within the organization. Like the NIST CSF, the ISO/IEC 27001:2013 standard is applicable to all organizations~\cite{iso_isoiec_nodate-1}.

\textit{Suitability and possible applications for SIEM implementation projects}: The annex of ISO/IEC 27001:2013 describes control objectives and controls. In this table of the standard, information on logging and monitoring is described under item A.12.4. In this context, the objective is understood to be the recording of events and the collection of evidence. Furthermore, the following four points are described on the subject of logging and monitoring, which provide more information~\cite{iso_isoiec_nodate-1}: 
\begin{itemize}
  \item Event Logging (A.12.4.1): Event logs should be created, retained, and reviewed on a regular basis. User activities, faults, exceptions and security events should be recorded.
  \item Protection of Log Information (A.12.4.2): The information should be protected from unauthorized manipulation and access.
  \item Administrator and Operator Logs (A.12.4.3): Activities of administrators and users on the systems are to be logged and these logs are to be protected and regularly reviewed.
  \item Clock Synchronization (A.12.4.4): The time of all relevant information processing systems within an organization must be synchronized with a single reference time.
\end{itemize}

The information from "Event Logging" and "Administrator and Operator Logs" could, for example, be taken into account when determining the log sources and use cases in the evaluation of a SIEM system. In addition, requirements for the SIEM can be derived from the references "Protection of Log Information" and "Clock Synchronization". In addition to the requirements from ISO/IEC 27001:2013, the ISO/IEC 27000 series also provides further information relating to ISMS.

\subsubsection{MITRE Att\&ck Framework}
The MITRE Att\&ck Framework is a knowledge base that contains tactics and techniques used by attackers. This database is managed and developed by the MITRE Corporation. The MITRE Corporation is an NPO and operates several government-funded research and development centers. One of these centers is responsible for cybersecurity research and development. MITRE's goal is to provide effective and practical tools and solutions for cybersecurity risk management. Furthermore, MITRE Corporation supports and advises NIST and the National Cybersecurity Center of Excellence (NCCoE) in the area of cybersecurity~\cite{mitre_corporation_mitre_nodate}.

MITRE's framework is a catalog consisting of techniques and tactics used by attackers. The framework contains 14 tactics, which in turn contain various techniques and sub-techniques. In addition to the 14 tactics, the framework contains a total of 191 techniques and 386 sub-techniques. Accordingly, the framework is referred to as a knowledge database, as companies can use it framework to learn which techniques are used by hackers when attacking companies. As an example, the tactic "Reconnaissance" describes techniques that are used to gather information, which in turn could be used to plan attacks. Further techniques such as "active scanning" and "phishing for information" are described. An organization can use the framework to deploy tools and solutions to trigger an alert when the corporate network is scanned or when multiple employees receive the similar phishing email. However, the framework does not describe which solutions and tools should be used to detect an attack. Nevertheless, the framework lists 40 data sources that could help a company detect certain attack techniques.

\textit{Suitability and possible applications for SIEM implementation projects}: The MITRE Att\&ck framework can be used as a source of information for defining use cases because of the tactics, techniques, and sub-techniques described. An obvious use case could be detecting phishing emails without relying on the user. In the MITRE Att\&ck framework, phishing is a technique with ID T1566. This technique is found under the Initial Access tactic, as attackers might try to obtain account access or place malware on users' devices. Usually, attackers send a victim an email, which in turn contains malicious attachments or links. The framework provides examples of such attacks for the techniques ('Procedure Example') and how they can be mitigated ('Mitigations') and detected ('Detection'). In this case, the framework also provides guidance on five measures that can be used to prevent phishing attacks. This information can be used to check which log sources could be used to detect such phishing attacks. Furthermore, Data Sources and their Data Components are mentioned, which should support the recognition of that kind of an attack. The question of which logs are required for a use case cannot be assessed in a generalized manner, since companies use different systems, applications and security tools. The MITRE Att\&ck Framework can therefore be used by companies for the definition of use cases.

\subsection{Elements of SIEM Implementation Procedure Model}
By analyzing existing procedure models, methods and frameworks, information could be collected which can be used for the development of our new procedure model. The artifacts of the new procedure model, which were identified in this analysis, are summarized compactly in the tables below.

\begin{table}[!htb]
  \centering
  \begin{tabular}{||l|p{5.2cm}||}
    \textbf{Approach} & \textbf{Description} \\
    Sequential & Applicable when requirements can be clearly defined from the beginning. \\
    Agile and iterative & Applicable when requirements cannot be clearly defined from the beginning and could change during the project. \\
  \end{tabular}
  \label{Table 3}
  \caption{Procedure Model Artifacts}
\end{table}

\begin{table}[!htb]
  \centering
  \begin{tabular}{||l|p{4.3cm}||}
    \textbf{Preocedure Phase} & \textbf{Tools \& Methods} \\
    Evaluation & Stakeholder analysis, requirements analysis, preference matrix, utility analysis, documentation, review \\
    Deployment \& Operation & Documentation, Review, Burn-Down-Chart, Scrum Board \\
  \end{tabular}
  \label{Table 4}
  \caption{Methods and Tools Artifacts}
\end{table}

\begin{table}[!htb]
  \centering
  \begin{tabular}{||l|p{5cm}||}
    \textbf{Preocedure Phase} & \textbf{Paragraph} \\
    Evaluation & NIST CSF: deriving requirements for a SIEM based on the References in the Detect Phase; ISO/IEC 27001: Deriving requirements for a SIEM based on logging and monitoring requirements; MITRE Att\&ck Framework: Derivation of use cases and the required required log sources \\
    Deployment & NIST CSF: Define the procedures and processes that will be followed in the event of an incident; ISO/IEC 27001: Consideration of the requirements Protection of Log Information and Clock Synchronization when Connecting Log Sources to the SIEM \\
    Operation & NIST CSF: Continuous monitoring of the security of the company through the SIEM; ISO/IEC 27001: Regular review of the collected information; MITRE Att\&ck Framework: Derivation of use cases and the required required log sources \\
  \end{tabular}
  \label{Table 5}
  \caption{Security Frameworks Artifacts}
\end{table}

\section{Development (4 pages)}
In this section, the new procedure model will be elaborated. It will start out with the basic elements of the model, and the result of this section will be the whole procedure model for SIEM implementation projects. 

\subsection{Implementation Prodecure Model}
\begin{figure}[!htb]
  \centering
  \includegraphics[width=80mm]{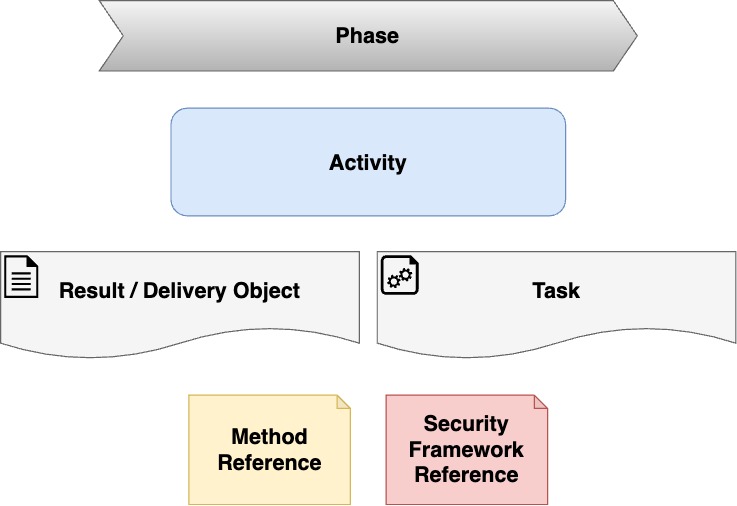}
  \caption{Procedure Model Elements} 
  \label{Figure 3}
\end{figure}

During development, six elements were used to model and represent the procedure model which are shown in Figure \ref{Figure 3}. The final model can be found here: \url{ https://github.com/correlatedsecurity/SPEED-SIEM-Use-Case-Framework} and a visualization is available here: \url{https://bit.ly/3C9UHaI}.

R2 (see Table \ref{Table 1}) states that the procedure model must consist of the phases Evaluation, Deployment and Operation. Since, in addition to the phases, activities and references, project management itself is also a component of a project, a layer was added to the model above the phases, which represents the magic triangle of project management. Therefore, the project management should monitor the time, scope and budget of the project across all phases. The "Scope" defines the desired result of the project, which should be achieved. "Time" represents the period of time in which the agreed project goal is to be achieved, and "Budget" shows the maximum costs that may be incurred to achieve the project goal within the desired time. Since, in addition to the phases, activities and references, project management itself is also a component of a project, a layer was added to the model above the phases, which represents the magic triangle of project management. The project management should monitor the time, scope and budget of the project across all phases. The "Scope" defines the desired result of the project, which should be achieved. "Time" represents the period of time in which the agreed project goal is to be achieved, and "Budget" shows the maximum costs that may be incurred to achieve the project goal within the desired time. In addition to the three phases of the process model, Figure 4 also shows the magic project management triangle~\cite{kuster_handbuch_2019}. 
\begin{figure}[!htb]
  \centering
  \includegraphics[width=80mm]{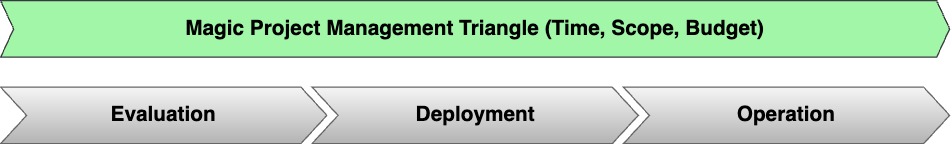}
  \caption{Phases incl. Magic Project Management Triangle}
  \label{Figure 4}
\end{figure}

\subsubsection{Evaluation}
The evaluation phase should help a company to evaluate the appropriate SIEM system that meets their requirements. But before these can be defined, the project team needs to know which people - or more generally spoken - which roles are allowed to define requirements for the new system as well as which people/roles will be affected by the implementation of the SIEM. A stakeholder analysis can be applied to identify these people (e.g., CISO, CTO, DPO). By analyzing the system landscape, it is possible to identify the core systems that are of primary importance during the initial deployment and whose logs should therefore be transmitted to the SIEM system on a mandatory basis. In addition, the findings of a system landscape analysis can be supportive at a later stage for the definition of the use cases when it comes to identifying the necessary logs that are required for mapping a specific use case (e.g., detection of phishing eMails). Consequently, the evaluation phase should start with an analysis of the initial situation, whereby both a stakeholder analysis and an analysis of the system landscape should be performed.

\begin{figure}[!htb]
  \centering
  \includegraphics[width=80mm]{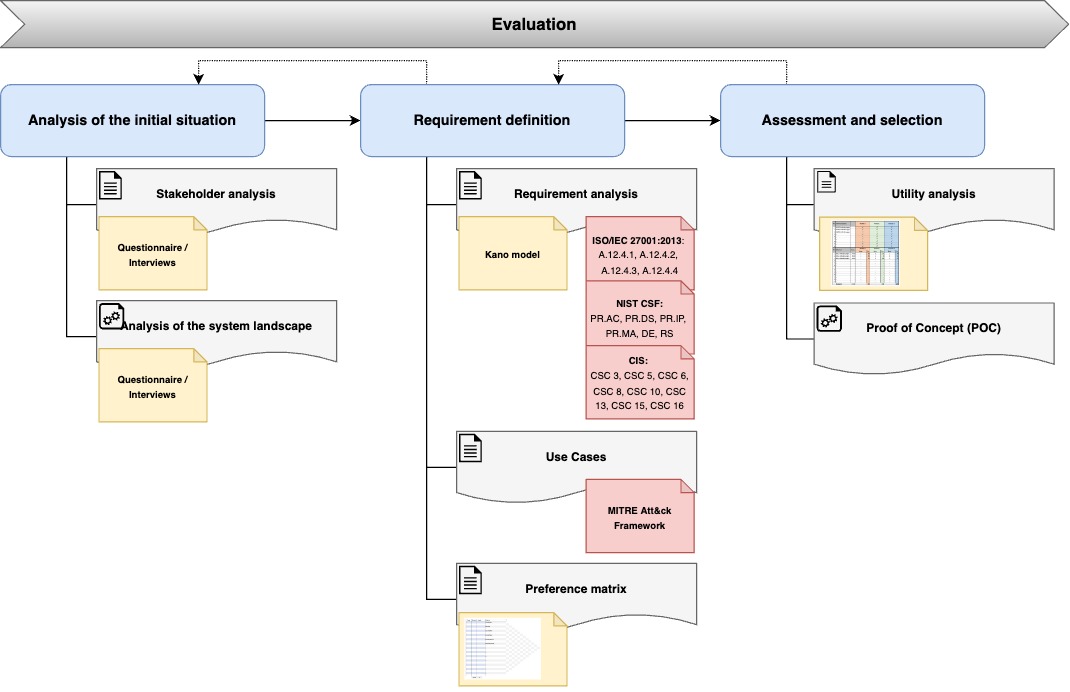}
  \caption{Draft of the Evaluation Phase} 
  \label{Figure 5}
\end{figure}

Upon analyzing the initial situation, the requirements which need to be fulfilled by a SIEM must be identified and defined. During the stakeholder analysis, it should have become clear which persons may define requirements for the system. Therefore, these persons should subsequently be surveyed using a questionnaire or an interview. The requirements from the requirement analysis can be defined according to the Kano model. The functional and non-functional requirements defined according to the Kano model are needed again in the next step for the evaluation and selection of the SIEM.

Once the requirements and the most important use cases have been defined, various SIEM systems should be evaluated on the basis of these findings. For this purpose, a preference matrix should first be created that ranks and weights the requirements for the system. Therefore, a preference matrix should be created in a meeting with all the necessary stakeholders. In this meeting, the stakeholders can exchange information with each other as well as discuss and decide together which requirements are most important for the company.

The assessment can be performed and visualized using a utility analysis. The product with the highest score is the most suitable SIEM system for the company based on the criteria previously defined by the stakeholders. For the evaluation of the system requirements, the developers' websites can be visited. Likewise, the developers or partners of the systems in question can be asked directly to obtain more information. Since, in addition to the functional and non-functional requirements, the use cases also play a role in the evaluation, research should be conducted to determine which SIEMs can map which use cases. 

\subsubsection{Deployment}
The installation and commissioning of the evaluated SIEM system is the overarching goal of the deployment phase. Before the system can be installed and put into operation, the contracts for obtaining the licenses or purchasing the software must be concluded. Kavanagh and Bussa~\cite{kavanagh_magic_2021} report in the SIEM Gartner Magic Quadrant Report that vendors outside their region often offer their product through partners or subsidiaries. Further, these partners and affiliates assist new customers in deploying the SIEM. Accordingly, licenses are often obtained through these partners rather than through the vendors themselves. In addition to the purchase, license and service contracts, local data protection laws should be taken into account, since personal data is transmitted, stored and analyzed in a SIEM.

\begin{figure}[!htb]
  \centering
  \includegraphics[width=80mm]{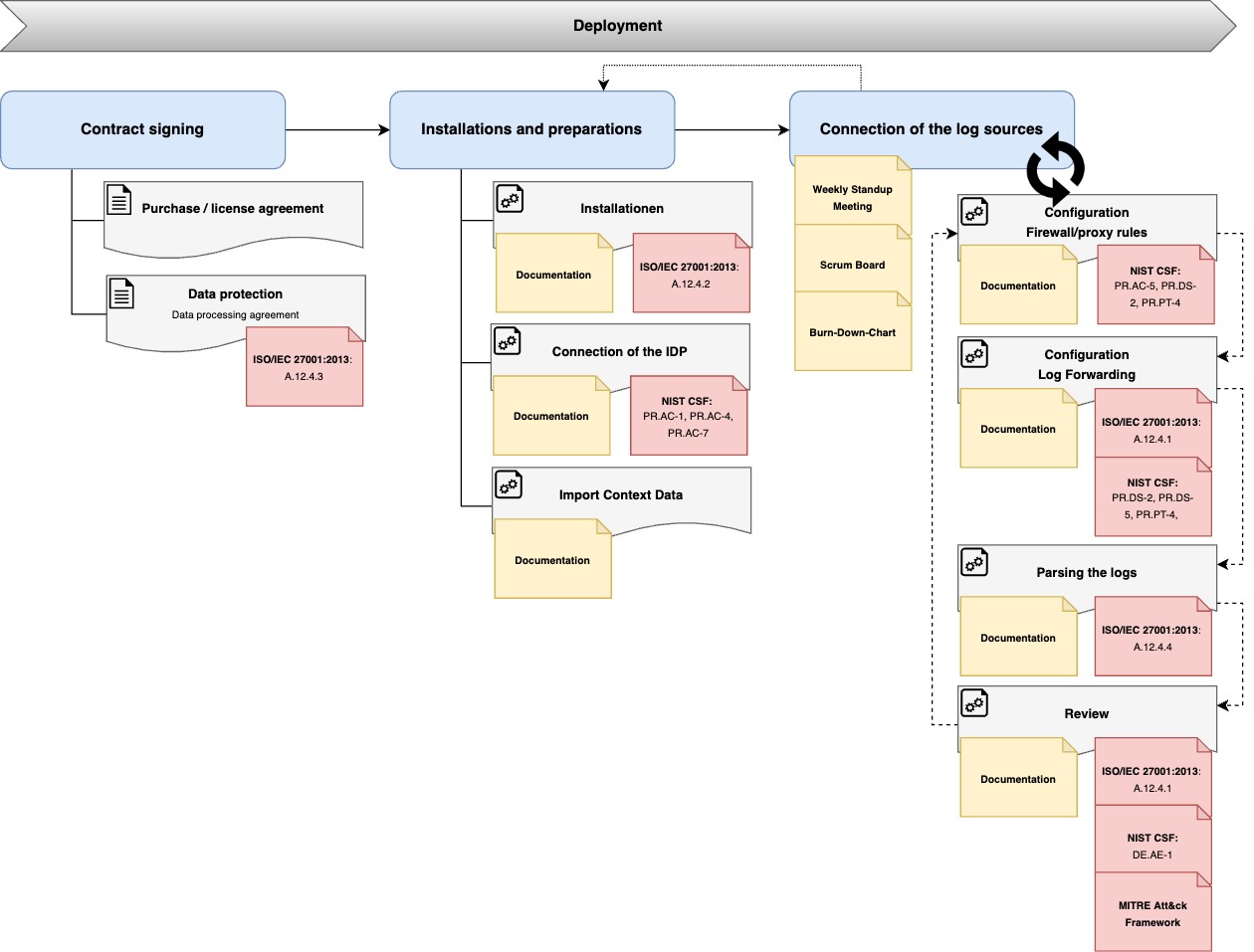}
  \caption{Draft of the Deployment Phase} 
  \label{Figure 6}
\end{figure}

Since the procedure model to be developed should be able to be used independently of manufacturer or product (see R4 in Table \ref{Table 1}), the focus of the deployment phase is on the connection of the log sources and not on which components have to be installed. Nevertheless, different installations must take place depending on the system. Even SaaS deployments require installations in the company's data center, depending on which logs should be transferred to the SIEM. In Azure Sentinel~\cite{mircosoft_log_nodate}, for example, agents are installed on the servers that send the logs to a gateway server, which forwards the logs received to the cloud-SIEM. Exabeam~\cite{exabeam_collectors_nodate} and Splunk~\cite{splunk_use_nodate} use collectors that forward the on-premise logs to the cloud. Therefore, the procedure model includes the task "Installation".

The activities "Contract signing" and "Installations and preparations" should be carried out sequentially, like the activities of the evaluation phase. In contrast, an agile form should be used for the connection of the log sources, for example with the help of Scrum Boards and/or a burn-down chart. Therefore, the integration of the log sources should take place in iterations, whereby an iteration contains four tasks, which have to be executed until a log source can be marked as implemented. 

\textit{1. Configuration firewall / proxy rules}: Before the logs are sent to the SIEM system from on-premise applications, servers, systems or appliances, it is necessary to open/enable the communication channels required for this purpose.

\textit{2. Configuration Log Forwardings}: After the ports are opened on the firewall, log forwarding can be configured. Unnecessary logs should be filtered out and not transferred to the SIEM. The configuration of log forwarding must be documented and made available to all authorized persons.

\textit{3. Parsing the logs}: As soon as the logs arrive in the SIEM, the parsing of the logs can be started. If the SIEM vendor already offers the appropriate parser, e.g. known and standardized logs, there is no need to create your own. Nevertheless, it is necessary to control how the logs are parsed and whether the required information is extracted. 

\textit{4. Review}: When the previous three steps are complete and the logs are parsed correctly, a review should take place. The goal of the review is to verify that all relevant logs are sent from a system to the SIEM and that the logs are parsed correctly. In addition, a completeness and quality check of the documentation should be performed.

In addition to the various activities, deliverables and tasks in the deployment phase, the model includes references to security frameworks and methods as defined in the requirements. These should be consolidated and taken into account in the individual activities.

\subsubsection{Operation}
When all defined log sources and use cases have been implemented through the various iterations in the deployment phase, the project can move into the final phase. The operations phase includes three activities. Here, the first two activities support the project team in handing over the SIEM to operations. The final activity looks at how the system should be operated, maintained and enhanced in the operational environment. 

\begin{figure}[!htb]
  \centering
  \includegraphics[width=80mm]{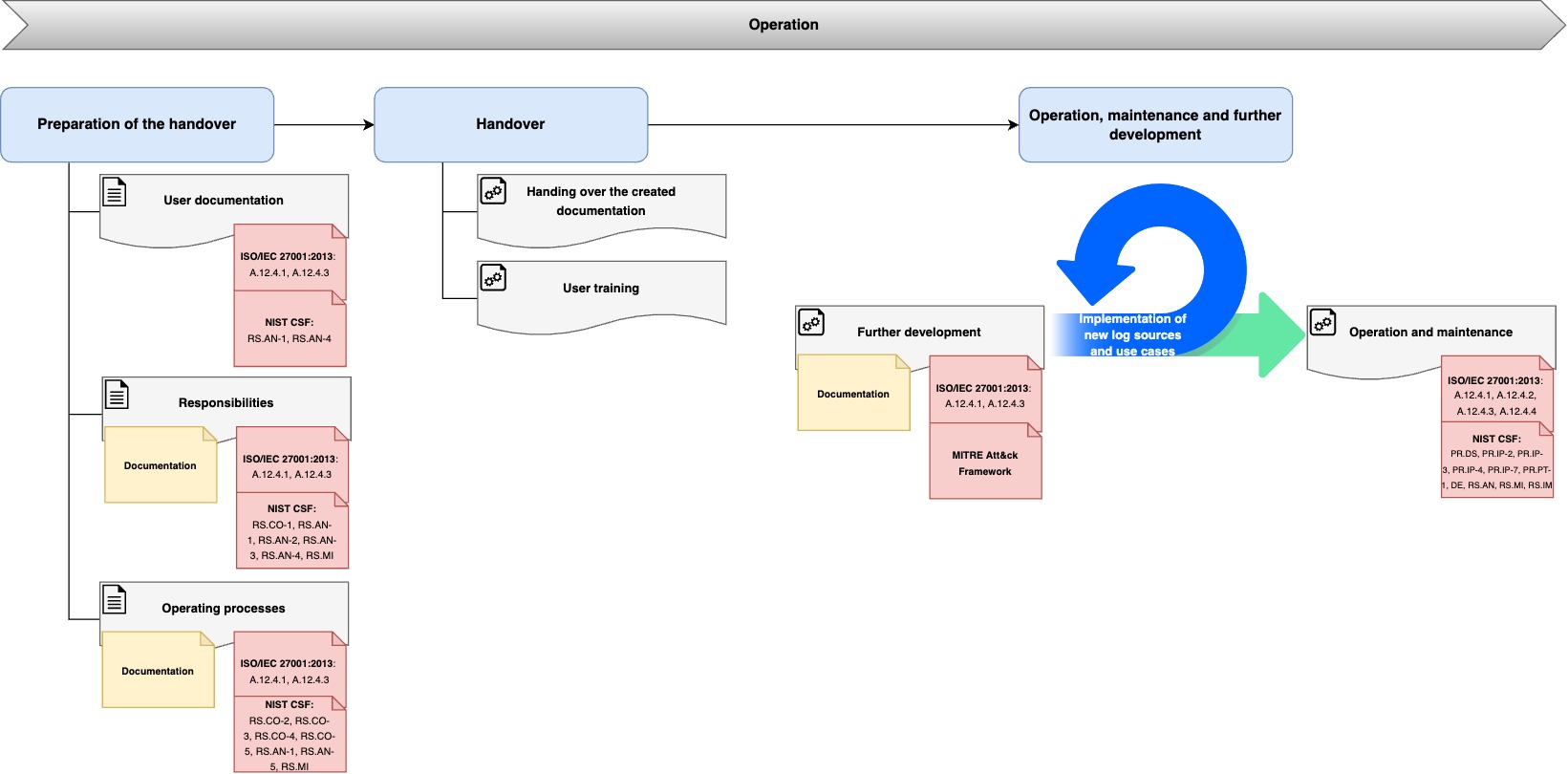}
  \caption{Draft of the Operation Phase} 
  \label{Figure 7}
\end{figure}

In the first step of the operating phase, the handover should be prepared. User documentation should be created, responsibilities defined and operational processes documented. User documentation supports the project team in training the defined operators of the system~\cite{scherb2023cyber}. In addition, the users can refer to this documentation at a later point in time and train new users in the same way. By defining and recording responsibilities, it is clearly agreed which persons are responsible for what. Furthermore, the processes for handling the SIEM system should be clarified and recorded. When defining responsibilities and operational processes, the NIST CSF should consider the respond ("RS") categories.

Once the documentation has been prepared, responsibilities and processes defined, the SIEM system can be handed over. This should involve introducing and training the users who will be working with the system in the future, before the system is effectively used by these individuals in their day-to-day work~\cite{scherb2023serious}. The activity also includes the task of handing over all artifacts and documents that were developed during the project. This ensures that the people working with the SIEM have access to all required information and documents.

When the first two activities of the operational phase are completed, the SIEM system can be handed over to operations. The SIEM system must be further developed by the operations team from this point on. The reason for this is that new applications and systems are constantly being introduced and old ones replaced in a company. As a result, every time applications or systems change new log sources have to be added. In addition, new use cases may need to be implemented due to new attack techniques, user needs, or changed and new corporate security policies. 

Consequently, the operational activities include the operation, maintenance and further development of the SIEM. The operation of a SIEM should therefore take place in an agile form, as visualized in the process model. With the help of an agile operational organization, incidents can be processed, new log sources can be integrated, and additional use cases can be implemented within a reasonable period of time. A variation of the suggested procedure would be a devops model, where the project and operations team would be closely integrated throughout the phases of the SIEM adoption.

\section{Validation}
In this section, the developed procedure model will be evaluated. It will start out with how the model was validated, and the result of this section will be the validated and adjusted procedure model for SIEM implementation projects.

\subsection{Artefact Evaluation}
After the development of the procedure model, a way had to be found to validate the new model. It was clear that the model should be evaluated by experts and not by people who had not yet carried out or supported a SIEM implementation project. A case study was conducted in which the newly developed procedure model (see section 5) was applied. The model was evaluated by two cybersecurity experts who have already carried out or accompanied various SIEM implementation projects. The aim of these interviews was to find out whether the model is applicable in this form and what can still be improved.

\textit{Requirements}: The interviews showed that the developed model fulfills all five requirements (see Table \ref{Table 1}) according to the experts. The model includes all three defined phases and the elements are based on existing project management methods that are used in practice (R1 and R2). Both experts also mentioned that the model was created from a high-level perspective, thus fulfilling the requirement that the model can be applied independently of the manufacturer or product (R3). Despite the high level view, important activities are included in the procedure model, which must be performed in a SIEM project. One reason for this is that the activities contain various results and tasks that provide users with information about which artefacts could be created. Likewise, the references to various methods and frameworks are helpful for users to understand how a result can be achieved or a task can be performed (R5). At this point, it must be mentioned that a procedure model should serve as a guide for a project and cannot be adopted 1:1. The reason for this is that corporations have defined internal guidelines and standards for projects, which must then also be taken into account. Since the procedure model considers the three phases of a SIEM implementation project from a high-level perspective, corporations have the option of incorporating additional activities, tasks and results that are specific to the product and/or corporation. Agility is maintained in the procedure model in the way that deliverables and tasks can be performed within the activities. Furthermore, the connection of the log sources and the operation must be iterative, which is why the requirements regarding agility are also met (R4).

\textit{Approach}: The experts are convinced that a purely sequential or agile model is generally not suitable for the introduction of software. The rationale for this is that in most projects there are phases that build on previous activities. The activities themselves are often carried out in projects through agile methods or an iterative process. The case study also showed that a sequential approach to evaluating the SIEM system was beneficial. The agile/iterative approach to deployment and operations also proved suitable. Furthermore, it was confirmed that the operation of a SIEM system must be carried out by an agile team. Otherwise, according to the experts, the system cannot be further developed on an ongoing basis, which in turn means that the effectively possible added value of a SIEM is not exploited.

\textit{Evaluation}: The interviews revealed that the most important components of an evaluation of a SIEM are present in the evaluation phase depicted. Nevertheless, there were a few suggestions for improvement from the experts. When analyzing the system landscape, in addition to the interviews, a document analysis would be very helpful if high-quality documentation is available. Many companies use a configuration management database (CMDB) in which the documentation of the various components of a corporation is stored. Both experts mentioned that systems should already be prioritized during the analysis of the system landscape. They said that the systems should be prioritized during this analysis on the basis of their relevance for the corporation. The background to this prioritization is that initial use cases can already be derived from such an analysis. In the case study, the corporation's systems were also prioritized for the subsequent definition of the use cases and log sources to be considered during the initial deployment. In addition, such prioritization can be used to determine which logs from which systems must actually be sent to a SIEM system.

Within the analysis of the initial situation, the IT strategy of the organization would still have to be examined, according to one expert. This should already take place before the stakeholder analysis and analysis of the system landscape. The IT strategy of an organization can already be used to gather initial findings. The analysis of the IT strategy makes it possible to find out what goals the IT organization wants to achieve by introducing a SIEM. Additionally, arguments for an investment in a SIEM system can be gathered, which have a direct impact on the achievement of the corporate strategy goals. Apart from that, one expert said that by analyzing the IT strategy, high level use cases and requirements can already be derived, which in turn could be considered as a basis for the next activity.

With regard to the security frameworks, the interviews showed that the process model contains references to the internationally known frameworks at the necessary points that are relevant in the SIEM context. Another framework would be the IT-Grundschutz from the Federal Office for Information Security of Germany~\cite{bundesamt_fur_sicherheit_in_der_informationstechnik_it-grundschutz_nodate}. An analysis of the BSI standards of the IT-Grundschutz (german for basic protection of IT Systems) has shown that only little information is provided with regard to SIEMs and log management, which, however, is already covered by the ISO/IEC 27001 references. As a result, it was decided that IT-Grundschutz would not be additionally included in the procedure model. In addition to the MITRE Att\&ck Framework, one expert also mentioned the SPEED SIEM Use Case Framework by Jurgen Visser for defining the use cases~\cite{visser_speed_2020}. This framework should support an organization in structuring, categorizing, defining, and describing use cases. In summary, the framework provides a template for how a company can organize and manage use cases in a structured way.

The case study showed that a proof of concept (POC) after a system was selected would have been beneficial to show the people who will later work with the system what the system can do and how their current workflow will be enriched by a SIEM system.

\textit{Deployment}: Both experts said that the deployment phase can be carried out in the manner described and includes the elements that must be present in a SIEM project from a high-level perspective. Nevertheless, there were two suggestions for improvement. In the contract signing activity, one expert found the topic of service level agreement (SLA) missing, which is particularly essential for cloud-based SIEM systems. The other expert noted that the architecture was not taken into account in the installation and preparation activity. According to the expert, the architecture should be a concept that includes the structure, the defined use cases and the log sources required for them and the roles and authorization management. For the creation of this concept, the artifacts created so far can be used and combined into one document. Based on this concept, the iterations for the log sources and the use cases integration in particular could be used in the further course of the deployment phase. The log sources and use cases could then be transferred from the concept to a Scrum Board, which the project team should use in the iterative process of deployment phase. With regard to the references to methods and security frameworks, the experts said that no important framework was missing or that ruther references to methods should be added to the model.

\textit{Operation}: According to the experts, the operation phase includes all activities, results and tasks that are required for the handover of the SIEM system. In addition, it was mentioned that the activity 'Operation, Maintenance and Further Development' shows that a SIEM is not only introduced once through an initial deployment, but must always be further developed and adapted form the corporation. In this context, it was criticized that the same element for the agile way of working should be used for the deployment phase as for the modeling in the operation phase. This would show more clearly that the iterations which are carried out during the connections of the log sources should be carried out again in the operational phase.

Regarding the Security Framework references, there were no comments from the experts that further requirements must be fulfilled to operate a SIEM. Instead, the creation of a RACI matrix (Responsible, Accountable, Consulted and Informed) was suggested, which would be beneficial for defining responsibilities. The expert justified this statement by saying that such a matrix clearly and transparently defines the tasks, responsibilities, and communication flow for decisions and changes in the corporation.

\subsection{Final Artefact}
Based on the inputs, extension and improvement suggestions, the process model was revised. Table~\ref{Table 6} lists the adjustments that were made to the procedure model based on the inputs and suggestions for improvement from the interviews with the two cybersecurity experts and the experiences from the case study.

\begin{table}[!htb]
  \centering
  \begin{tabular}{||l|p{6.3cm}||}
    \textbf{Phase} & \textbf{Changes} \\
    Evaluation & Adjustment for the method reference questionnaire/interview. \\
    & Adjustment for the delivery object Use Cases. \\
    & Adjustment of the task Proof of Concept (POC). \\
    & New task Analysis of IT Strategy added. \\
    & New delivery object roadmap and outlook added. \\
    & \\
    Deployment & New result Service Level Agreement (SLA) added. \\
    & New delivery object architecture added. \\
    & Task connection of the IDP removed. \\
    & Adjustments of the activity connection of the log sources. \\
    & \\
    Operation & Adjustments of the delivery object User documentation \\
    & Adjustments of the delivery object responsibilities \\
    & Adjustments of the delivery object business processes \\
    & Adjustments of the task further development \\
  \end{tabular}
  \label{Table 6}
  \caption{List of modifications of the developed procedure model}
\end{table}

\begin{figure}[!htb]
  \centering
  \includegraphics[width=80mm]{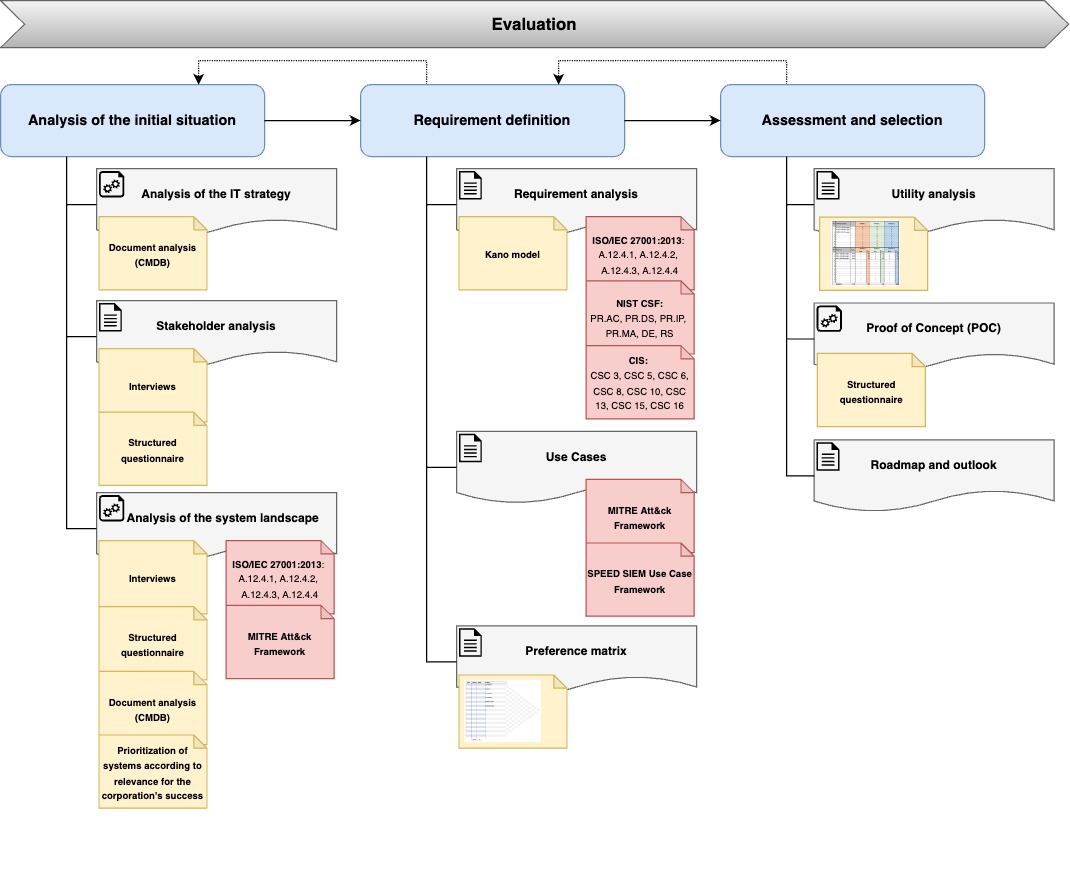}
  \caption{Evaluation Phase} 
  \label{Figure 8}
\end{figure}

\begin{figure}[!htb]
  \centering
  \includegraphics[width=80mm]{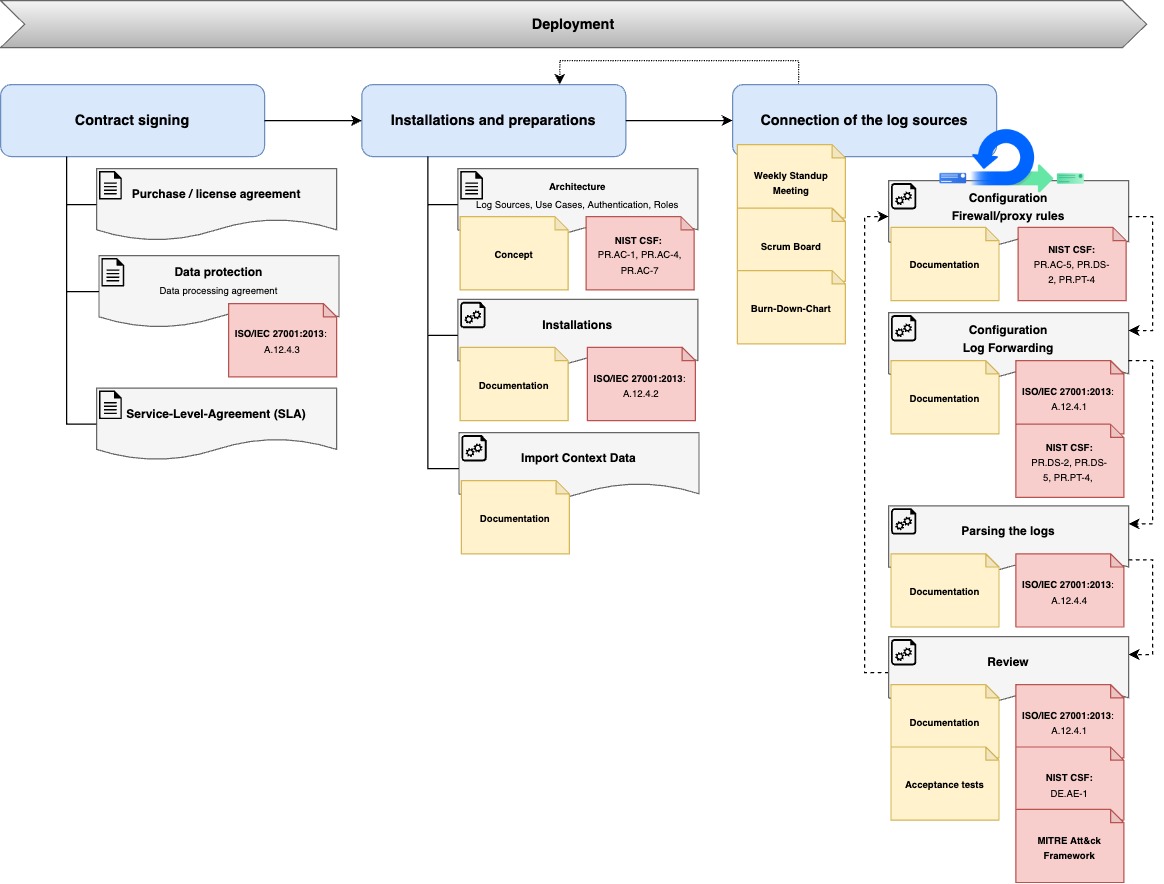}
  \caption{Deployment Phase} 
  \label{Figure 9}
\end{figure}

\begin{figure}[!htb]
  \centering
  \includegraphics[width=80mm]{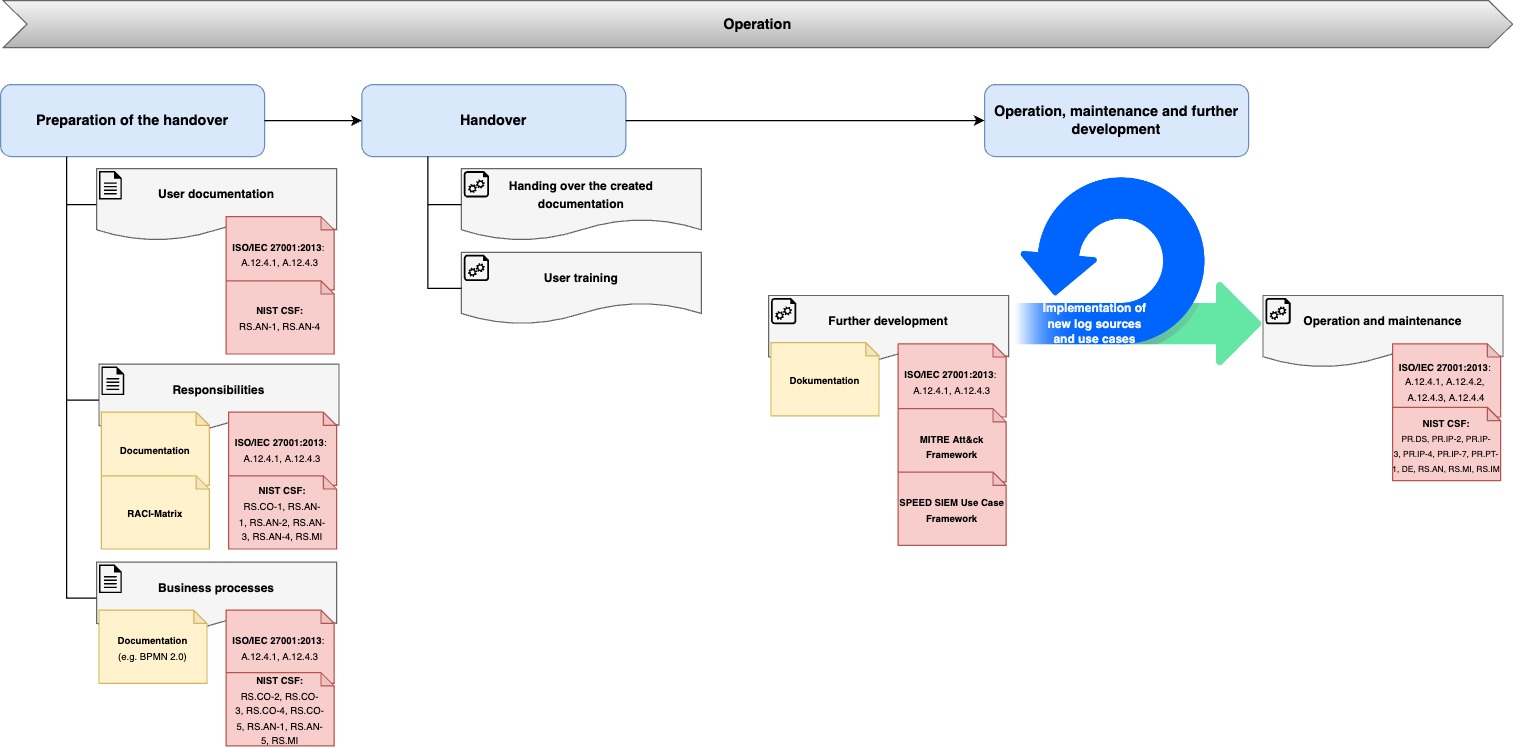}
  \caption{Deployment Phase} 
  \label{Figure 10}
\end{figure}

Figure \ref{Figure 8} shows the evaluation phase, Figure \ref{Figure 9} the deployment phase, and Figure \ref{Figure 10} the operation phase of the developed, evaluated, and adjusted procedure model for the implementation of a SIEM system in corporations. In addition, the complete procedure model can be viewed and downloaded as a PDF file via the link in the footnote \footnote[1]{https://bit.ly/3C9UHaI}.

\section{Conclusion and Outlook}
Within the framework of this research, four lead questions were addressed, which were defined in the introduction (see Introduction). The chapter 'Problem Awareness' addressed the first two lead questions. By conducting a literature review, the basic functionalities of log management and SIEM systems were identified. In addition, the current state of SIEMs was considered and it was described how the market has developed in recent years. In the chapter 'Analysis' classical procedure models and security frameworks were investigated. It was researched whether the waterfall model, the V-Model XT and the Scrum approach can be used for SIEM implementation projects. Furthermore, the three frameworks - NIST CSF, ISO/IEC 27001:2013 and the MITRE Att\&ck Framework - were examined with regard to suitable elements for the procedure model. The result of this research is a developed procedure model (chapter 'Development). In the interviews as part of the chapter 'Validation', selected cybersecurity experts confirmed the practical suitability of the procedure model. 

In addition, the model was reflected by means of a practical case study at a Swiss industrial company. The development of the artifact and the practical case study took place at the same time in many parts. The following further measure is desired for the future: the new process model should be applied in several new companies to test to prove its general use in practice. Through systematic supervision and evaluation, the model could be further improved and, if necessary, expanded. Currently, the model contains a manageable number of references to three security frameworks. It would be conceivable that in a next stage further frameworks, standards, and best practices could be identified, which could be could be used in a SIEM implementation project. In order to keep the model clear and simple, an additional diagram could be created in which the various references would be categorized and cataloged.

Our developed procedure model provides the following benefits: 1) It is based on existing project management methods and can therefore be applied in a simple and straightforward manner; 2) it holistically covers the phases 'Evaluation', 'Deployment' and 'Operation'; 3) it is neutral and can be used independently of any vendor or product; 4) it is based on agile methods and approaches, which makes it flexible and adaptable; 5) as a special feature, it contains references to further methods and/or security frameworks.

This work contributes to solving the problem that so far there is no procedure model for the implementation of SIEM systems for companies. Using Henver's DSR approach, we created a methodologically grounded artifact that is publicly available. Our research is intended to produce a benefit for practitioners who are responsible for a SIEM implementation in their organization. Likewise, the model is at disposal for critical appraisals, further developments and adaptations of the research and practitioner community.




\bibliographystyle{IEEEtran}
\bibliography{IEEEabrv, Paper}

\begin{thebibliography}{10}
\providecommand{\url}[1]{#1}
\csname url@samestyle\endcsname
\providecommand{\newblock}{\relax}
\providecommand{\bibinfo}[2]{#2}
\providecommand{\BIBentrySTDinterwordspacing}{\spaceskip=0pt\relax}
\providecommand{\BIBentryALTinterwordstretchfactor}{4}
\providecommand{\BIBentryALTinterwordspacing}{\spaceskip=\fontdimen2\font plus
\BIBentryALTinterwordstretchfactor\fontdimen3\font minus
  \fontdimen4\font\relax}
\providecommand{\BIBforeignlanguage}[2]{{%
\expandafter\ifx\csname l@#1\endcsname\relax
\typeout{** WARNING: IEEEtran.bst: No hyphenation pattern has been}%
\typeout{** loaded for the language `#1'. Using the pattern for}%
\typeout{** the default language instead.}%
\else
\language=\csname l@#1\endcsname
\fi
#2}}
\providecommand{\BIBdecl}{\relax}
\BIBdecl

\bibitem{bygrave_cyber_2022}
L.~A. Bygrave, ``Cyber {Resilience} versus {Cybersecurity} as {Legal}
  {Aspiration},'' in \emph{2022 14th {International} {Conference} on {Cyber}
  {Conflict}: {Keep} {Moving}! ({CyCon})}, vol. 700, May 2022, pp. 27--43,
  iSSN: 2325-5374.

\bibitem{national_institute_of_standards_and_technology_framework_2018}
\BIBentryALTinterwordspacing
{National Institute of Standards and Technology},
  ``\BIBforeignlanguage{en}{Framework for {Improving} {Critical}
  {Infrastructure} {Cybersecurity}, {Version} 1.1},'' National Institute of
  Standards and Technology, Gaithersburg, MD, Tech. Rep. NIST CSWP 04162018,
  Apr. 2018. [Online]. Available:
  \url{http://nvlpubs.nist.gov/nistpubs/CSWP/NIST.CSWP.04162018.pdf}
\BIBentrySTDinterwordspacing

\bibitem{chuvakin_complete_2021}
\BIBentryALTinterwordspacing
A.~Chuvakin, ``\BIBforeignlanguage{en}{The {Complete} {Guide} to {Log} and
  {Event} {Management}},'' 2021. [Online]. Available:
  \url{https://www.microfocus.com/media/white-paper/the-complete-guide-to-log-and-event-management-wp.pdf}
\BIBentrySTDinterwordspacing

\bibitem{mokalled_guidelines_2019}
\BIBentryALTinterwordspacing
H.~Mokalled, R.~Catelli, V.~Casola, D.~Debertol, E.~Meda, and R.~Zunino,
  ``\BIBforeignlanguage{en}{The {Guidelines} to {Adopt} an {Applicable} {SIEM}
  {Solution}},'' \emph{\BIBforeignlanguage{en}{Journal of Information
  Security}}, vol.~11, no.~1, pp. 46--70, Dec. 2019, number: 1 Publisher:
  Scientific Research Publishing. [Online]. Available:
  \url{http://www.scirp.org/Journal/Paperabs.aspx?paperid=97094}
\BIBentrySTDinterwordspacing

\bibitem{cybersecurity_workforce_isc_nodate}
\BIBentryALTinterwordspacing
{Cybersecurity Workforce}, ``({ISC})² 2021 {Cybersecurity} {Workforce}
  {Study}.'' [Online]. Available:
  \url{https://www.isc2.org:443/Research/Workforce-Study}
\BIBentrySTDinterwordspacing

\bibitem{huang_comprehensive_2016}
\BIBentryALTinterwordspacing
T.~Huang and K.~Yasuda, ``Comprehensive review of literature survey articles on
  {ERP},'' \emph{Business Process Management Journal}, vol.~22, no.~1, pp.
  2--32, Jan. 2016, publisher: Emerald Group Publishing Limited. [Online].
  Available: \url{https://doi.org/10.1108/BPMJ-12-2014-0122}
\BIBentrySTDinterwordspacing

\bibitem{kurbel_erp_2013}
\BIBentryALTinterwordspacing
K.~E. Kurbel, ``\BIBforeignlanguage{en}{{ERP} {System} {Implementation}},'' in
  \emph{\BIBforeignlanguage{en}{Enterprise {Resource} {Planning} and {Supply}
  {Chain} {Management}: {Functions}, {Business} {Processes} and {Software} for
  {Manufacturing} {Companies}}}, ser. Progress in {IS}, K.~E. Kurbel, Ed.\hskip
  1em plus 0.5em minus 0.4em\relax Berlin, Heidelberg: Springer, 2013, pp.
  159--188. [Online]. Available:
  \url{https://doi.org/10.1007/978-3-642-31573-2_6}
\BIBentrySTDinterwordspacing

\bibitem{kronbichler_comparison_2009}
S.~Kronbichler, H.~Ostermann, and R.~Staudinger, ``A {COMPARISON} {OF}
  {ERP}-{SUCCESS} {MEASUREMENT} {APPROACHES},'' \emph{JISTEM Journal of
  Information Systems and Technology Management; Vol 7, No 2 (2010); 281-310},
  vol.~7, Dec. 2009.

\bibitem{hevner_design_2004}
A.~Hevner, A.~R, S.~March, S.~T, {Park}, J.~Park, {Ram}, and {Sudha}, ``Design
  {Science} in {Information} {Systems} {Research},'' \emph{Management
  Information Systems Quarterly}, vol.~28, p.~75, Mar. 2004.

\bibitem{kuechler_theory_2008}
\BIBentryALTinterwordspacing
B.~Kuechler and V.~Vaishnavi, ``On theory development in design science
  research: anatomy of a research project,'' \emph{European Journal of
  Information Systems}, vol.~17, no.~5, pp. 489--504, Oct. 2008, publisher:
  Taylor \& Francis \_eprint: https://doi.org/10.1057/ejis.2008.40. [Online].
  Available: \url{https://doi.org/10.1057/ejis.2008.40}
\BIBentrySTDinterwordspacing

\bibitem{kent_guide_2006}
\BIBentryALTinterwordspacing
K.~Kent and M.~P. Souppaya, ``\BIBforeignlanguage{en}{Guide to computer
  security log management},'' National Institute of Standards and Technology,
  Gaithersburg, MD, Tech. Rep. NIST SP 800-92, 2006, edition: 0. [Online].
  Available:
  \url{https://nvlpubs.nist.gov/nistpubs/Legacy/SP/nistspecialpublication800-92.pdf}
\BIBentrySTDinterwordspacing

\bibitem{sahoo_syslog_2019}
P.~Sahoo, R.~Chottray, G.~Jena, and S.~Pattnaiak, ``Syslog a {Promising}
  {Solution} to {Log} {Management},'' May 2019.

\bibitem{miloslavskaya_analysis_2018}
N.~Miloslavskaya, ``\BIBforeignlanguage{en}{Analysis of {SIEM} {Systems} and
  {Their} {Usage} in {Security} {Operations} and {Security} {Intelligence}
  {Centers}},'' in \emph{\BIBforeignlanguage{en}{Biologically {Inspired}
  {Cognitive} {Architectures} ({BICA}) for {Young} {Scientists}}}, ser.
  Advances in {Intelligent} {Systems} and {Computing}, A.~V. Samsonovich and
  V.~V. Klimov, Eds.\hskip 1em plus 0.5em minus 0.4em\relax Cham: Springer
  International Publishing, 2018, pp. 282--288.

\bibitem{vielberth_security_2021}
M.~Vielberth, ``Security {Information} and {Event} {Management} ({SIEM}),'' in
  \emph{Encyclopedia of {Cryptography}, {Security} and {Privacy}}, Mar. 2021.

\bibitem{yelevin_use_2021}
\BIBentryALTinterwordspacing
{Yelevin} and {Batamig}, ``\BIBforeignlanguage{en-us}{Use entities to classify
  and analyze data in {Microsoft} {Sentinel}},'' Nov. 2021. [Online].
  Available: \url{https://docs.microsoft.com/en-us/azure/sentinel/entities}
\BIBentrySTDinterwordspacing

\bibitem{salitin_role_2018}
M.~Salitin and A.~Zolait, ``The role of {User} {Entity} {Behavior} {Analytics}
  to detect network attacks in real time,'' Nov. 2018, pp. 1--5.

\bibitem{exabeam_exabeam_2021}
\BIBentryALTinterwordspacing
{Exabeam}, ``Exabeam {Use} {Case} {Series}: {Contextualization},'' Jul. 2021.
  [Online]. Available:
  \url{https://community.exabeam.com/s/article/Exabeam-Use-Case-Series-Contextualization}
\BIBentrySTDinterwordspacing

\bibitem{voigt_ueba_2018}
\BIBentryALTinterwordspacing
L.~Voigt, ``\BIBforeignlanguage{en-US}{{UEBA} and {Why} {It} {Should} {Be}
  {Part} of {Your} {Incident} {Response}},'' Aug. 2018, section: UEBA.
  [Online]. Available:
  \url{https://www.exabeam.com/ueba/ueba-uba-siem-incident-response/}
\BIBentrySTDinterwordspacing

\bibitem{shahid_anomaly_2021}
N.~Shahid and M.~Ali~Shah, ``{ANOMALY} {DETECTION} {IN} {SYSTEM} {LOGS} {IN}
  {THE} {SPHERE} {OF} {DIGITAL} {ECONOMY},'' in \emph{Competitive {Advantage}
  in the {Digital} {Economy} ({CADE} 2021)}, vol. 2021, Jun. 2021, pp.
  185--190.

\bibitem{scherb2018resolution}
C.~Scherb, D.~Grewe, M.~Wagner, and C.~Tschudin, ``Resolution strategies for
  networking the iot at the edge via named functions,'' in \emph{2018 15th IEEE
  Annual Consumer Communications \& Networking Conference (CCNC)}.\hskip 1em
  plus 0.5em minus 0.4em\relax IEEE, 2018, pp. 1--6.

\bibitem{grewe2018network}
D.~Grewe, C.~Marxer, C.~Scherb, M.~Wagner, and C.~Tschudin, ``A network stack
  for computation-centric vehicular networking,'' in \emph{Proceedings of the
  5th ACM Conference on Information-Centric Networking}, 2018, pp. 208--209.

\bibitem{gonzalez-granadillo_security_2021}
\BIBentryALTinterwordspacing
G.~González-Granadillo, S.~González-Zarzosa, and R.~Diaz, ``Security
  {Information} and {Event} {Management} ({SIEM}): {Analysis}, {Trends}, and
  {Usage} in {Critical} {Infrastructures},'' \emph{Sensors (Basel,
  Switzerland)}, vol.~21, no.~14, p. 4759, Jul. 2021. [Online]. Available:
  \url{https://www.ncbi.nlm.nih.gov/pmc/articles/PMC8309804/}
\BIBentrySTDinterwordspacing

\bibitem{sukma_analysis_2019}
N.~Sukma, W.~Srisawat, P.~Sa-nga ngam, and A.~Leelasantitham, ``An {Analysis}
  of {Log} {Management} {Practices} to reduce {IT} {Operational} {Costs}
  {Using} {Big} {Data} {Analytics},'' in \emph{2019 4th {Technology}
  {Innovation} {Management} and {Engineering} {Science} {International}
  {Conference} ({TIMES}-{iCON})}, Dec. 2019, pp. 1--5.

\bibitem{kavanagh_magic_2021}
\BIBentryALTinterwordspacing
K.~Kavanagh, T.~Bussa, and J.~Collins, ``Magic {Quadrant} for {Security}
  {Information} and {Event} {Management},'' Gartner, Tech. Rep., Jun. 2021.
  [Online]. Available:
  \url{https://www.gartner.com/doc/reprints?id=1-28NL22KV&ct=220107&st=sb}
\BIBentrySTDinterwordspacing

\bibitem{alienvault_siem_2022}
\BIBentryALTinterwordspacing
{AlienVault}, ``\BIBforeignlanguage{EN}{The {SIEM} {Evaluator}'s {Guide}},''
  AlienVault, Tech. Rep., 2022. [Online]. Available:
  \url{https://cdn-cybersecurity.att.com/docs/guides/The-SIEM-Evaluators-Guide.pdf}
\BIBentrySTDinterwordspacing

\bibitem{splunk_leitfaden_2018}
{Splunk}, ``\BIBforeignlanguage{de}{Der {Leitfaden} für {SIEM}-{Käufer}},''
  2018.

\bibitem{filkins_evaluators_2018}
\BIBentryALTinterwordspacing
B.~Filkins, ``\BIBforeignlanguage{EN}{An {Evaluator}’s {Guide} to {NextGen}
  {SIEM}},'' SANS-Institut, Tech. Rep., Dec. 2018. [Online]. Available:
  \url{https://logrhythm.com/resources/independent-white-papers/sans-evaluators-guide-to-nextgen-siem/}
\BIBentrySTDinterwordspacing

\bibitem{safarzadeh_novel_2019}
M.~Safarzadeh, H.~Gharaee, and A.~H. Panahi, ``\BIBforeignlanguage{en}{A
  {Novel} and {Comprehensive} {Evaluation} {Methodology} for {SIEM}},'' in
  \emph{\BIBforeignlanguage{en}{Information {Security} {Practice} and
  {Experience}}}, S.-H. Heng and J.~Lopez, Eds.\hskip 1em plus 0.5em minus
  0.4em\relax Cham: Springer International Publishing, 2019, pp. 476--488.

\bibitem{grigorof_azure_2021}
\BIBentryALTinterwordspacing
A.~Grigorof, M.~Mocanu, and J.~Shaw-Young, ``\BIBforeignlanguage{en}{Azure
  {Sentinel} {Deployment} {Best} {Practices}},'' 2021. [Online]. Available:
  \url{https://i.crn.com/sites/default/files/ckfinderimages/userfiles/images/crn/custom/2021/BlueVoyant_CloseUp_Whitepaper_Microsoft_Azure_Sentinel_Deployment_Q3_2021.pdf}
\BIBentrySTDinterwordspacing

\bibitem{holik_deployment_2015}
F.~Holik, J.~Horalek, S.~Neradova, S.~Zitta, and O.~Marik, ``The deployment of
  {Security} {Information} and {Event} {Management} in cloud infrastructure,''
  in \emph{2015 25th {International} {Conference} {Radioelektronika}
  ({RADIOELEKTRONIKA})}, Apr. 2015, pp. 399--404.

\bibitem{asprion2023agile}
P.~M. Asprion, C.~Giovanoli, C.~Scherb, and S.~Bhat, ``Agile management in
  cybersecurity,'' \emph{Proceedings of Society}, vol.~93, pp. 21--32, 2023.

\bibitem{broy_vorgehensmodelle_2021}
\BIBentryALTinterwordspacing
M.~Broy and M.~Kuhrmann, ``\BIBforeignlanguage{de}{Vorgehensmodelle in der
  {Softwareentwicklung}},'' in \emph{\BIBforeignlanguage{de}{Einführung in die
  {Softwaretechnik}}}, ser. Xpert.press, M.~Broy and M.~Kuhrmann, Eds.\hskip
  1em plus 0.5em minus 0.4em\relax Berlin, Heidelberg: Springer, 2021, pp.
  83--124. [Online]. Available:
  \url{https://doi.org/10.1007/978-3-662-50263-1_3}
\BIBentrySTDinterwordspacing

\bibitem{heil_vorgehensmodelle_2012}
\BIBentryALTinterwordspacing
A.~Heil, ``\BIBforeignlanguage{de}{Vorgehensmodelle},'' in
  \emph{\BIBforeignlanguage{de}{Anwendungsentwicklung für {Intelligente}
  {Umgebungen} im {Web} {Engineering}}}, A.~Heil, Ed.\hskip 1em plus 0.5em
  minus 0.4em\relax Wiesbaden: Springer Fachmedien, 2012, pp. 37--77. [Online].
  Available: \url{https://doi.org/10.1007/978-3-8348-2551-3_3}
\BIBentrySTDinterwordspacing

\bibitem{leyh_passende_2019}
\BIBentryALTinterwordspacing
C.~Leyh, ``\BIBforeignlanguage{de}{Passende {ERP}-{Systeme} auswählen und
  einführen},'' \emph{\BIBforeignlanguage{de}{Controlling \& Management
  Review}}, vol.~63, no.~5, pp. 52--57, Jul. 2019. [Online]. Available:
  \url{https://doi.org/10.1007/s12176-019-0027-4}
\BIBentrySTDinterwordspacing

\bibitem{sandhaus_hybride_2014}
\BIBentryALTinterwordspacing
G.~Sandhaus, B.~Berg, and P.~Knott, ``\BIBforeignlanguage{de}{Hybride
  {Vorgehensmodelle}},'' in \emph{\BIBforeignlanguage{de}{Hybride
  {Softwareentwicklung}: {Das} {Beste} aus klassischen und agilen {Methoden} in
  einem {Modell} vereint}}, ser. Xpert.press, B.~Berg, P.~Knott, and
  G.~Sandhaus, Eds.\hskip 1em plus 0.5em minus 0.4em\relax Berlin, Heidelberg:
  Springer, 2014, pp. 53--62. [Online]. Available:
  \url{https://doi.org/10.1007/978-3-642-55064-5_4}
\BIBentrySTDinterwordspacing

\bibitem{schatten_vorgehensmodelle_2010}
\BIBentryALTinterwordspacing
A.~Schatten, M.~Demolsky, D.~Winkler, S.~Biffl, E.~Gostischa-Franta, and
  T.~Östreicher, ``\BIBforeignlanguage{de}{Vorgehensmodelle},'' in
  \emph{\BIBforeignlanguage{de}{Best {Practice} {Software}-{Engineering}:
  {Eine} praxiserprobte {Zusammenstellung} von komponentenorientierten
  {Konzepten}, {Methoden} und {Werkzeugen}}}, A.~Schatten, M.~Demolsky,
  D.~Winkler, S.~Biffl, E.~Gostischa-Franta, and T.~Östreicher, Eds.\hskip 1em
  plus 0.5em minus 0.4em\relax Heidelberg: Spektrum Akademischer Verlag, 2010,
  pp. 47--69. [Online]. Available:
  \url{https://doi.org/10.1007/978-3-8274-2487-7_3}
\BIBentrySTDinterwordspacing

\bibitem{vivenzio_vorgehensmodelle_2013}
\BIBentryALTinterwordspacing
A.~Vivenzio and D.~Vivenzio, ``\BIBforeignlanguage{de}{Vorgehensmodelle in der
  {Softwareentwicklung}},'' in \emph{\BIBforeignlanguage{de}{Testmanagement bei
  {SAP}-{Projekten}: {Erfolgreich} {Planen} • {Steuern} • {Reporten} bei
  der {Einführung} von {SAP}-{Banking}}}, A.~Vivenzio and D.~Vivenzio,
  Eds.\hskip 1em plus 0.5em minus 0.4em\relax Wiesbaden: Springer Fachmedien,
  2013, pp. 5--10. [Online]. Available:
  \url{https://doi.org/10.1007/978-3-8348-2142-3_2}
\BIBentrySTDinterwordspacing

\bibitem{royce_managing_1970}
\BIBentryALTinterwordspacing
D.~W.~W. Royce, ``\BIBforeignlanguage{en}{Managing the development of large
  software systems},'' p.~11, 1970. [Online]. Available:
  \url{https://www.praxisframework.org/files/royce1970.pdf}
\BIBentrySTDinterwordspacing

\bibitem{gessler_projektphasen_2016}
M.~Gessler and R.~Kaestner, ``Projektphasen,'' Sep. 2016, pp. 349--366.

\bibitem{kneuper_geschichtliche_2018}
R.~Kneuper, \emph{Die geschichtliche {Entwicklung} des {V}-{Modells}}, Nov.
  2018.

\bibitem{angermeier_v-modell_2019}
\BIBentryALTinterwordspacing
D.~Angermeier, C.~Bartelt, O.~Bauer, G.~Beneken, K.~Bergner, U.~Birowicz,
  T.~Bliß, C.~Breitenstrom, N.~Cordes, D.~Cruz, P.~Dohrmann, J.~Friedrich,
  M.~Gnatz, U.~Hammerschall, I.~Hidvegi-Barstorfer, H.~Hummel, D.~Israel,
  T.~Klingenberg, K.~Klugseder, I.~Küffer, M.~Kuhrmann, W.~Kranz, M.~Kranz,
  H.-J. Meinhardt, M.~Meisinger, S.~Mittrach, H.-J. Neußer, D.~Niebuhr,
  K.~Plögert, D.~Rauh, A.~Rausch, T.~Rittel, W.~Rösch, E.~Saas, J.~Schramm,
  M.~Sihling, T.~Ternité, S.~Vogel, B.~Weber, and M.~Wittmann,
  ``\BIBforeignlanguage{DE}{V-{Modell}® {XT}},'' Jan. 2019. [Online].
  Available:
  \url{http://ftp.tu-clausthal.de/pub/institute/informatik/v-modell-xt/Releases/2.3/V-Modell-XT-Gesamt.pdf}
\BIBentrySTDinterwordspacing

\bibitem{kleuker_vorgehensmodelle_2018}
\BIBentryALTinterwordspacing
S.~Kleuker, ``\BIBforeignlanguage{de}{Vorgehensmodelle},'' in
  \emph{\BIBforeignlanguage{de}{Grundkurs {Software}-{Engineering} mit {UML}:
  {Der} pragmatische {Weg} zu erfolgreichen {Softwareprojekten}}}, S.~Kleuker,
  Ed.\hskip 1em plus 0.5em minus 0.4em\relax Wiesbaden: Springer Fachmedien,
  2018, pp. 25--54. [Online]. Available:
  \url{https://doi.org/10.1007/978-3-658-19969-2_3}
\BIBentrySTDinterwordspacing

\bibitem{volland_scrum-framework_2021}
\BIBentryALTinterwordspacing
M.~F. Volland, \emph{\BIBforeignlanguage{de}{Das {Scrum}-{Framework} in
  {Großunternehmen} – {Entscheidung} und {Implementation}: {Eine}
  {Fallstudie} eines multinationalen {Automobilkonzerns}}}.\hskip 1em plus
  0.5em minus 0.4em\relax Wiesbaden: Springer Fachmedien, 2021. [Online].
  Available: \url{https://link.springer.com/10.1007/978-3-658-35001-7}
\BIBentrySTDinterwordspacing

\bibitem{takeuchi_new_1986}
\BIBentryALTinterwordspacing
H.~Takeuchi and I.~Nonaka, ``The {New} {New} {Product} {Development} {Game},''
  \emph{Harvard Business Review}, Jan. 1986, section: Product development.
  [Online]. Available:
  \url{https://hbr.org/1986/01/the-new-new-product-development-game}
\BIBentrySTDinterwordspacing

\bibitem{lucht_theorie_2019}
\BIBentryALTinterwordspacing
D.~Lucht, \emph{\BIBforeignlanguage{de}{Theorie und {Management} komplexer
  {Projekte}}}.\hskip 1em plus 0.5em minus 0.4em\relax Wiesbaden: Springer
  Fachmedien, 2019. [Online]. Available:
  \url{http://link.springer.com/10.1007/978-3-658-14476-0}
\BIBentrySTDinterwordspacing

\bibitem{nist_cybersecurity_2013}
\BIBentryALTinterwordspacing
{NIST}, ``\BIBforeignlanguage{en}{Cybersecurity {Framework}},''
  \emph{\BIBforeignlanguage{en}{NIST}}, Nov. 2013, last Modified:
  2022-09-09T08:52-04:00. [Online]. Available:
  \url{https://www.nist.gov/cyberframework}
\BIBentrySTDinterwordspacing

\bibitem{iso_isoiec_nodate}
\BIBentryALTinterwordspacing
{ISO}, ``\BIBforeignlanguage{en}{{ISO}/{IEC} 27001 — {Information} security
  management}.'' [Online]. Available:
  \url{https://www.iso.org/isoiec-27001-information-security.html}
\BIBentrySTDinterwordspacing

\bibitem{mitre_corporation_mitre_nodate}
\BIBentryALTinterwordspacing
{MITRE Corporation}, ``{MITRE} {ATT}\&{CK}®.'' [Online]. Available:
  \url{https://attack.mitre.org/}
\BIBentrySTDinterwordspacing

\bibitem{iso_isoiec_nodate-1}
\BIBentryALTinterwordspacing
{ISO}, ``\BIBforeignlanguage{en}{{ISO}/{IEC} 27001:2013}.'' [Online].
  Available:
  \url{https://www.iso.org/cms/render/live/en/sites/isoorg/contents/data/standard/05/45/54534.html}
\BIBentrySTDinterwordspacing

\bibitem{kuster_handbuch_2019}
\BIBentryALTinterwordspacing
J.~Kuster, C.~Bachmann, E.~Huber, M.~Hubmann, R.~Lippmann, E.~Schneider,
  P.~Schneider, U.~Witschi, and R.~Wüst,
  \emph{\BIBforeignlanguage{en}{Handbuch {Projektmanagement}}}, 4th~ed.\hskip
  1em plus 0.5em minus 0.4em\relax Springer Gabler Berlin, Heidelberg, 2019.
  [Online]. Available: \url{https://doi.org/10.1007/978-3-662-57878-0}
\BIBentrySTDinterwordspacing

\bibitem{mircosoft_log_nodate}
\BIBentryALTinterwordspacing
{Mircosoft}, ``\BIBforeignlanguage{en-us}{Log {Analytics} agent overview -
  {Azure} {Monitor}}.'' [Online]. Available:
  \url{https://docs.microsoft.com/en-us/azure/azure-monitor/agents/log-analytics-agent}
\BIBentrySTDinterwordspacing

\bibitem{exabeam_collectors_nodate}
\BIBentryALTinterwordspacing
{Exabeam}, ``\BIBforeignlanguage{en-US}{Collectors}.'' [Online]. Available:
  \url{https://docs.exabeam.com/collectors/}
\BIBentrySTDinterwordspacing

\bibitem{splunk_use_nodate}
\BIBentryALTinterwordspacing
{Splunk}, ``Use forwarders to get data into {Splunk} {Cloud} {Platform} -
  {Splunk} {Documentation}.'' [Online]. Available:
  \url{https://docs.splunk.com/Documentation/SplunkCloud/8.2.2202/Data/UsingforwardingagentsCloud}
\BIBentrySTDinterwordspacing

\bibitem{scherb2023cyber}
C.~Scherb, L.~B. Heitz, F.~Grimberg, H.~Grieder, and M.~Maurer, ``A cyber
  attack simulation for teaching cybersecurity,'' \emph{EPiC Series in
  Computing}, vol.~93, pp. 129--140, 2023.

\bibitem{scherb2023serious}
------, ``A serious game for simulating cyberattacks to teach cybersecurity,''
  \emph{arXiv preprint arXiv:2305.03062}, 2023.

\bibitem{bundesamt_fur_sicherheit_in_der_informationstechnik_it-grundschutz_nodate}
\BIBentryALTinterwordspacing
{Bundesamt für Sicherheit in der Informationstechnik},
  ``\BIBforeignlanguage{de}{{IT}-{Grundschutz}}.'' [Online]. Available:
  \url{https://www.bsi.bund.de/DE/Themen/Unternehmen-und-Organisationen/Standards-und-Zertifizierung/IT-Grundschutz/it-grundschutz.html?nn=128656}
\BIBentrySTDinterwordspacing

\bibitem{visser_speed_2020}
\BIBentryALTinterwordspacing
J.~Visser, ``\BIBforeignlanguage{en}{{SPEED} {SIEM} {Use} {Case}
  {Framework}},'' Feb. 2020. [Online]. Available:
  \url{https://github.com/correlatedsecurity/SPEED-SIEM-Use-Case-Framework}
\BIBentrySTDinterwordspacing

\end{thebibliography}


\vfill


\end{document}